\documentclass[12pt]{JHEP}
\usepackage{amsfonts} %to get blackboard bold, bold etc
\usepackage{amsbsy} %likewise

\newcommand{\C}[1]{{\mathcal #1}}
\newcommand{\beq}{\begin{equation}}
\newcommand{\eeq}{\end{equation}}
\newcommand{\bea}{\begin{eqnarray}}
\newcommand{\eea}{\end{eqnarray}}
\newcommand{\Tr}{{\hbox{Tr}\,}}
\newcommand{\comm}[2]{\left[#1,#2\right]}
\newcommand{\absval}[1]{{\left\vert#1\right\vert}}
\newcommand{\diag}[1]{{\hbox{diag}\,#1}}
\newcommand{\thetafn}[1]{{\,\theta\!#1}}
\newcommand{\deltafn}[1]{{\,\delta\!#1}}
\newcommand{\expect}[1]{{<#1>}}

\newcommand{\half}{{1\over 2}}

\newcommand{\nn}{\nonumber}

\title{The Convergence of Yang-Mills Integrals}

\author{Peter Austing\\ Department of Physics, University of Oxford \\
Theoretical Physics,\\
1 Keble Road,\\
 Oxford OX1 3NP, UK\\
E-mail: \email{p.austing@physics.ox.ac.uk}}
\author{John F. Wheater\\ Department of Physics, University of Oxford \\
Theoretical Physics,\\
1 Keble Road,\\
 Oxford OX1 3NP, UK\\
E-mail: \email{j.wheater@physics.ox.ac.uk}}

\preprint{OUTP-01-01P, \hepth{0101071}}
\abstract{ We prove that $SU(N)$ bosonic Yang-Mills matrix integrals
are convergent for dimension (number of matrices) $D\ge D_c$. It is
already known that $D_c=5$ for $N=2$; we prove that $D_c=4$ for $N=3$
and that  $D_c=3$ for $N\ge 4$. These results are consistent with the 
numerical evaluations of the integrals by Krauth and Staudacher.
}

\begin{document}

\section{Introduction}
The discovery of D-branes and the realization of their importance
in string theory and M-theory has led to a number of exciting conjectures
relating these very complicated theories to (at least technically) much simpler M(atrix) theories \cite{Banks:1997vh} and the IKKT model for the type
IIB superstring \cite{Ishibashi:1997xs} (for  reviews
see \cite{Taylor:1999qk,Aoki:1998bq}).
This has generated  renewed interest in supersymmetric
Yang-Mills quantum mechanics \cite{Sethi:1997pa,Yi:1997eg,Kac:1999av,Hoppe:1999xg,Staudacher:2000gx} which is 
obtained by the the dimensional 
reduction of $D=10$ supersymmetric
Yang-Mills gauge theory (SSYM) to 1 
remaining space-time dimension.
Further dimensional reduction to 0 dimensions leads to 
the supersymmetric Yang-Mills  matrix integrals which are an important 
component in calculating the Witten index for the quantum mechanics. Both the 
quantum mechanics and the matrix integrals exist for any dimension
$D$ in which the original SSYM exists, ie $D=3$, 4,  6, and 10.
Dropping the supersymmetry requirement leads to the bosonic Yang-Mills matrix 
integrals which exist for any $D$ and are the main subject of
 this paper.

A couple of years ago Moore, Nekrasov and Shatashvili \cite{Moore:1998et}
found a way of calculating the partition function  by deforming the Yang-Mills matrix integrals to a cohomological 
theory. Their method does not allow the calculation of arbitrary 
correlation functions in the original Yang-Mills picture where it is not known 
how to do exact calculations unless the gauge group is $SU(2)$ 
\cite{Suyama:1998ig}. However it is possible to do numerical  calculations
provided the gauge group is not too big. The numerically calculated partition
functions for small $N=3$, 4, and 5 \cite{Krauth:1998xh,Krauth:1998yu}
agree with the cohomological calculations in \cite{Moore:1998et}. Some
correlation functions and eigenvalue distributions have also been obtained
\cite{Krauth:1998yu,Krauth:1999qw}; in the case of $D=4$ where the fermions
can be integrated numerically by Monte Carlo many correlation functions
have been found for values of $N$ up to 48 \cite{Ambjorn:2000bf}.
 In the course of their work Krauth and Staudacher \cite{Krauth:1998yu} also investigated the properties of purely bosonic Yang-Mills
integrals (these are defined below).   It had been believed that the flat directions in the action would  cause these integrals to diverge (their supersymmetric
cousins being saved by the Pfaffian arising from the integration of the
fermions which  vanishes along the flat directions).   Simple analytic calculations of the partition functions
 in the case of $SU(2)$, and delicate numerical computations for 
$SU(3)$, $SU(4)$ and $SU(5)$  (and subsequently for $N$ up to 256 \cite{Hotta:1998en}
and 
other gauge groups as well \cite{Krauth:2000bv})
showed that this is not necessarily the case.  Unfortunately up to now
an analytic demonstration of the convergence of these integrals 
for $SU(N>2)$ has been lacking. Our purpose here is to provide such a 
demonstration.

The bosonic Yang-Mills partition function for  gauge group
$G$   in  $D$ ``space-time'' dimensions \footnote{In the bosonic case there
is no requirement of supersymmetry to restrict $D$ so it is possible to consider $D$ as a continuous variable by analytic continuation; however in this paper it is to be taken strictly as an integer.} 
dimensionally reduced to 0  is given by
\beq \C Z_{D,G}=\prod_{\mu=1}^D\int_{-\infty}^\infty dX_\mu\exp\left(\sum_{\mu>\nu}
\Tr \comm{X_\mu}{X_\nu}^2\right) \label{P1}\eeq
where the  matrices $\{X_\mu, \mu=1,\ldots D\}$, which are traceless and hermitian,
take values  in the Lie algebra of $G$ and can be written
\beq X_\mu=\sum_{a=1}^g X_\mu^a t^a.\label{P2}\eeq
The $\{t^a, a=1,\ldots g\}$ are the generators in the fundamental representation satisfying
\beq \Tr t^a t^b=2\delta^{ab}\label{P3}\eeq
and we shall  use $l$ to denote the rank of the Lie algebra. 
 In this
paper we will restrict ourselves to the groups $SU(N)$.  The measure and the
integrand in \ref{P1} are then invariant under the $SU(N)$ gauge symmetry
\beq X_\mu\to U^\dagger X_\mu U,\qquad U\in SU(N)\label{I1}\eeq
and the $SO(D)$ symmetry
\beq X_\mu\to \sum_\nu Q_{\mu\nu}X_\nu,\qquad Q\in SO(D).\label{I2}\eeq
 We will prove that $\C Z_{D,SU(N)}$ is
 convergent for dimension $D\ge D_c$ and divergent for $D<D_c$.
It is already known from exact calculation that $D_c=5$ for $N=2$
(although we will show that our methods reproduce this  almost
trivially). We prove that $D_c=4$ for $N=3$
and that  $D_c=3$ for $N\ge 4$.

The body of this  paper is concerned with establishing which integrals
\emph{converge}. In section 2 we set up our procedure 
and establish some results that are useful in every case. Section 3 deals with $SU(2)$, section 4 with $SU(3)$, and section 5 with $SU(N>3)$. In Appendix
A we show which integrals \emph{diverge}.  We conclude with a brief discussion
of the implications of our results for the supersymmetric theories and other
gauge groups.

\section{Preliminaries}

The dangerous regions which might cause the integral \ref{P1}  to diverge
are where all the commutators almost vanish but the magnitude of 
$X_\mu$  goes to infinity. Hence
 we let 
\beq X_\mu=Rx_\mu,\quad \Tr x_\mu x_\mu=1 \label{P4}\eeq
where, as from now on, we  use the summation convention for Greek indices.
Then we have
\beq \C Z_{D, G}=\int_0^\infty R^{Dg-1} \C X_{D, G}(R) dR\label{P5}\eeq
where
\beq \C X_{D, G}(R)=\prod_{\nu=1}^D\int dx_\nu\,\deltafn{\left(1-\Tr x_\mu x_\mu\right)}
\exp\left(-R^4S\right) \label{P6}\eeq
and
\bea S&=&-\half\Tr \comm{x_\mu}{x_\nu}\comm{x_\mu}{x_\nu}\nn\\
&=&\half\sum_{i,j,\mu,\nu}\absval{\comm{x_\mu}{x_\nu}_{i,j}}^2.\label{P6.0}\eea
We note that for any finite $R$ the integral $\C X_{D, G}(R)$ is bounded by
a constant (since every term in the argument of the exponential is negative
semi-definite)  and therefore if for large $R$
\bea\C X_{D, G}(R)&<&\frac{const}{R^\alpha},\qquad\hbox{where}\:\alpha> Dg,\label{P6.1}\eea
then the partition function $\C Z_{D, G}$ is finite. Our tactic for proving
convergence of $\C Z_{D, G}$ is therefore to find a bound of the form
\ref{P6.1} on $\C X_{D, G}(R)$.

Now we split the integration region in \ref{P6} into two
\bea \C R_1:&& S
<(R^{-(2-\eta)})^2\nn\\
\C R_2:&& S
\ge(R^{-(2-\eta)})^2\label{P7}\eea
where $\eta$ is small but positive.  We see immediately that the contribution
to $\C I_{D, G}(R)$ from $\C R_2$ is bounded by $A_1\exp(-R^{2\eta})$ (we will
use the capital letters
$A$, $B$ and $C$ to denote constants throughout this paper)
 and thus automatically satisfies \ref{P6.1}. Thus we can
confine our efforts to the contribution from $\C R_1$ in which we replace
the exponential function by unity to get the bound
\beq \C X_{D, G}(R)<A_1\exp(-R^{2\eta})+\C I_{D, G}(R) 
 \label{P7.1}\eeq
where 
\beq\C I_{D, G}(R) = \prod_{\nu=1}^D\int_{\C R_1} dx_\nu
\deltafn{\left(1-\Tr x_\mu x_\mu\right)}.
 \label{P7.2}\eeq 
Since $S$ is a sum of squares it follows 
that the region $\C R_1'$ defined
by
\beq \absval{\comm{x_\mu}{x_\nu}_{ij}}<R^{-(2-\eta)},\quad
\forall{\mu,\nu,i,j}\label{P8}\eeq
is larger than the region $\C R_1$ which we can therefore replace 
in \ref{P7.1} by $\C R_1'$.

Now we utilise the $SU(N)$ symmetry  to diagonalise $x_1$ which we may
therefore write as $x_1=\hbox{ diag}(\lambda_{1},\ldots\lambda_{N})$.
The constraint \ref{P8} then becomes
\beq\absval{(\lambda_{i}-\lambda_{j})(x_\nu)_{ij}}<R^{-(2-\eta)}\label{P9}\eeq
This immediately leads to the generic case  which is when the eigenvalues
of $x_1$ are not degenerate 
\beq \absval{\lambda_{i}-\lambda_{j}}>\epsilon\label{P9.1}\eeq
where $\epsilon$ is a constant which we may choose but will always
be finite. Then
all the off-diagonal elements of $\{x_\nu, \nu=2,\ldots D\}$ are bounded by
\ref{P9} and 
$\C I^{}_{D, G}(R)$ (\ref{P7.2})
 has a contribution $\C I^{gen}_{D, G}(R)$ which is bounded 
\beq \C I^{gen}_{D, G}(R) < A_2 R^{-(2-\eta)(D-1)(g-l)}\label{P10}\eeq
where the constant $A_2$ comes from the integral over all the diagonal
elements (and does of course depend on $\epsilon$).
Note that once the off-diagonal elements are bounded by \ref{P9} and \ref{P9.1}
all commutators are constrained to be $O(R^{-(2-\eta)})$ although
 the coefficient may be more than 1. Enforcing \ref{P8} implies  constraints
 on the diagonal elements which lower $A_2$ in \ref{P10} but 
does not affect the 
power of $R$.

We note that the case when all eigenvalues are widely separated corresponds
to the ``perturbation'' expansion of \cite{Hotta:1998en}; using \ref{P6.1} and
\ref{P10} leads to the criterion $N>D/(D-2)$ for convergence that is given
in \cite{Hotta:1998en}. It is fortuitous
that this gives the correct conditions for convergence of $\C Z_{D,G}$
because as we shall see $\C I^{gen}_{D,G}$ is not usually the most divergent
contribution.

\section{$SU(2)$}

In the case of $SU(2)$ each matrix $x_\mu$ has eigenvalues 
$\lambda_\mu$ and $-\lambda_\mu$. However the constraint in \ref{P4}
implies that
\beq 2\lambda_\mu\lambda_\mu=1\label{SU2-1}\eeq
and it follows that there must always be one matrix with eigenvalues
of magnitude at least $(2D)^{-\half}$. We can  choose this matrix
to be $x_1$ and hence by taking $\epsilon < 2(2D)^{-\half}$ we always
have the generic case. Hence
\beq  \C X_{D,SU(2)}(R)<A_1\exp(-R^{2\eta})+A_2 R^{-2(2-\eta)(D-1)}
\label{SU2-2}\eeq
and it follows immediately that $\C Z_{D,SU(2)}$ is finite for $D\ge 5$.
We show in Appendix A that $\C Z_{D,SU(2)}$ is divergent for smaller $D$. 
 Of course
these results are  well known because  $\C Z_{D,SU(2)}$ can be
 calculated exactly.

\section{$SU(3)$}

From the constraint in \ref{P4} it follows that there must always be one matrix with an eigenvalue
of magnitude at least $(3D)^{-\half}$; as before we choose this matrix 
to be $x_1$. When all the eigenvalues of $x_1$ are separated by at least 
$\epsilon$ (\ref{P9.1}) we get the generic contribution 
\ref{P10} to the integral. However  now we have the new possibility that two of the eigenvalues
of $x_1$ become degenerate 
  (it is not possible for
all three eigenvalues to become degenerate provided we choose $\epsilon<\half(3D)^{-\half}$)
in which case the condition \ref{P9.1} does
not apply and we have to proceed more carefully; $\C I_{D, SU(3)}(R)$
is made up of a piece where \ref{P9.1} applies to all eigenvalues
plus a new piece $\C I^{deg}_{D, SU(3)}(R)$ from the region of
integration where \ref{P9.1} is not satisfied by all eigenvalues,
\beq \C I_{D, SU(3)}(R)=\C I^{gen}_{D, SU(3)}(R)+\C I^{deg}_{D, SU(3)}(R).\label{SU3-1}\eeq

 First we write
\beq x_1=\frac{\rho_1}{\sqrt 3}\left(\begin{array}{ccc}
					1& 0& 0\\
					0& 1& 0\\
					0& 0& -2\end{array}\right)
+\left(\begin{array}{ccc}
					\tilde x_1& & 0\\
					& & 0\\
					0& 0& 0\end{array}\right)
\label{SU3-2}\eeq
where $\tilde x_1=\diag{(\xi,-\xi)}$; we know that $\absval{2\rho_1/{\sqrt 3}}>(3D)^{-\half}$ and,
 because we are
just interested in the case of degenerate eigenvalues,
 $2\absval{\xi}<\epsilon$.  Because only two of the three eigenvalues
of $x_1$ are degenerate the constraint \ref{P9} bounds some of the off-diagonal
elements of $x_{2,\ldots D}$ so we can write these matrices in the form
\beq x_\nu=\frac{\rho_\nu}{\sqrt 3}\left(\begin{array}{ccc}
					1& 0& 0\\
					0& 1& 0\\
					0& 0& -2\end{array}\right)
+\left(\begin{array}{ccc}
					\tilde x_\nu& & 0\\
					& & 0\\
					0& 0& 0\end{array}\right)
+\left(\begin{array}{ccc}
					0&0&O(R^{-(2-\eta)})\\
					0&0&O(R^{-(2-\eta)})\\
			O(R^{-(2-\eta)})&O(R^{-(2-\eta)})&0\end{array}\right)
\label{SU3-3}\eeq
where $\tilde x_\nu$ is a $2\times 2$ traceless hermitian matrix (ie it lives
in the $su(2)$ sub-algebra of the $su(3)$ algebra inhabited by $x_\nu$), and by $O(R^{-(2-\eta)})$ we mean that the elements are bounded by \ref{P9}. At 
this stage the off-diagonal elements in the third row and column of the
 $x_{2,\ldots D}$
are innocuous and can be integrated out  to get 
\bea \C I^{deg}_{D,SU(3)}(R)& <B_1 &R^{-(2-\eta)(4(D-1)-2)}
\left(\int_{\absval{\rho_1}>(4D)^{-\half}}\prod_{\mu=1}^D d\rho_\mu\right)\left( \int_{\widetilde{\C R}} d\xi\prod_{\nu=2}^D d\tilde x_\nu\right)
(2\xi)^2\nn\\&&\times
\Omega_{SU(2)}({\sqrt 3}\rho_1-\xi)^2({\sqrt 3}\rho_1+\xi)^2
\thetafn{\left(1-\left(\Tr{\tilde x_\alpha\tilde x_\alpha} +2\rho_\alpha\rho_\alpha\right)\right)}\nn\\&&\times
\thetafn{\left(\left(\Tr{\tilde x_\alpha\tilde x_\alpha} +2\rho_\alpha\rho_\alpha\right)-\left(1-
4\left(\epsilon R^{(2-\eta)}\right)^{-2}\right)\right)}
\label{SU3-4}\eea
where $\theta$ denotes the step function, 
 we have included the Vandermonde determinant for $x_1$, and $\Omega_{SU(2)}$
is the volume of $SU(2)$.  The region $\widetilde{\C R}$ is defined by $\absval{\xi}<\epsilon/2$ and
\beq \absval{\comm{\tilde x_\mu}{\tilde x_\nu}_{ij}}<R^{-(2-\eta)},\quad
\forall{\mu,\nu,i,j}.\label{SU3-5}\eeq
Within the region of integration we can bound the factors
$(3\rho_1\pm \xi)^2$ by a constant  and then,  because we are looking for 
an upper bound, we can drop the constraints $\absval{\xi}<\epsilon/2$
and $\absval{\rho_1}>(4D)^{-\half}$. Doing the $\rho_\mu $ integrals
we are left with
\beq\C I^{deg}_{D,SU(3)}(R) <B_2 R^{-(2-\eta)4(D-1)}\C F_{D,SU(2)}\label{SU3-6}\eeq
where
\bea \C F_{D,SU(2)}&=&\left(\int_{\widetilde{\C R}} d\xi\prod_{\nu=2}^D d\tilde x_\nu\right)
(2\xi)^2\Omega_{SU(2)}\thetafn{\left(1-\Tr{\tilde x_\mu\tilde x_\mu}\right)}\nn\\
&=&(2-\eta) R^{-(2-\eta)}\int_0^Rdu\, u^{1-\eta}
\left(\int_{\widetilde{\C R}} \prod_{\nu=1}^D d\tilde x_\nu\right)
\deltafn{\left(\left[\frac{u}{R}\right]^{2-\eta}-\Tr{\tilde x_\mu\tilde x_\mu}\right)}.
\label{SU3-7}\eea
Making the rescaling $\tilde  x_\mu=\tilde y_\mu[u/R]^{1-\eta/2}$
we find that
\bea \C F_{D,SU(2)}&=& R^{-(2-\eta)3D/2}\int_0^R u^{3D(1-\eta/2)-1}\C I_{D,SU(2)}(u)
\, du.\label{SU3-8}\eea
As $R\to\infty $ we use the results from section 3 to bound 
 the remaining integral
giving
\bea   \C F_{D,SU(2)}&<&  B_3 R^{-(2-\eta)3D/2},\quad  D\ge 5\nn\\
	&&B_4 R^{-(2-\eta)3D/2}\log R,\quad D=4\nn\\
	&&B_5 R^{-(2-\eta)(3D-1)/2},\quad D=3\label{SU3-9}\eea
and so we find that
\bea  \C I^{deg}_{D,SU(3)}(R) &<& B_6
\left(\frac{1}{ R^{2-\eta}}\right)^{4(D-1)+(3D-\delta_{D,3})/2}
\left(\log R\right)^{\delta_{D,4}}.\label{SU3-10}\eea
From \ref{P10} we know that 
\beq \C I^{gen}_{D, SU(3)}(R) < B_7 R^{-(2-\eta)6(D-1)}.\label{SU3-11}\eeq
Applying the criterion \ref{P6.1} we see that both
 $\C I^{deg}_{D, SU(3)}(R)$ and $\C I^{gen}_{D, SU(3)}(R)$ 
 make a finite contribution to 
$ \C Z_{D,SU(3)}$ only for $D\ge 4$.  It is straightforward to check that $ \C Z_{D=3,SU(3)}$ diverges 
(see appendix). Thus we conclude that $ \C Z_{D,SU(3)}$
is finite for $D\ge 4$ only.

\section{$N>3$}

The argument for higher $N$ has the same structure as for $SU(3)$.
 Again there is a generic
contribution which is bounded as shown in \ref{P10} and a degenerate
contribution. As usual $x_1$ can be chosen so that at least one of its
eigenvalues has magnitude $(ND)^{-\half}$ or more.  The difference is that $x_1$ can have not only two degenerate
but $2,3,\ldots N-1$ and two or more sets of them. To deal with this we 
note that whenever $x_1$ has some configuration of exactly degenerate
eigenvalues there is a sub-algebra of $su(N)$ which commutes with $x_1$.
(As we saw in section 4 for the case of $SU(3)$ the only possibility is to
have two degenerate eigenvalues and the sub-algebra is $su(2)$.)  Suppose
that $x_1$ has $K$ sets of degenerate eigenvalues with
degeneracies $\{N_1,\ldots N_K\}$;  then the sub-algebra which commutes with $x_1$ is
\beq H=\bigoplus_{k=1}^{K} su(N_k)\label{SUN-1}\eeq
%We will denote the number of elements of $H$ by $h$ and the rank
%by $m$.

Now let $\bar t^a $  denote those generators of $G$ that do
not lie
in  $H$, and decompose $x_\mu$ into
\bea
 x_1&=&\diag{\left(\rho^1_1 I^{N_1},\ldots \rho^K_1 I^{N_K},
\sigma^1_1\ldots \sigma^M_1 \right)}+
\diag{\left(\tilde x^1_1,\ldots \tilde x^K_1,
0\ldots 0\right)}\nn\\
 x_{\mu>1}&=&\diag{\left(\rho^1_\mu I^{N_1},\ldots \rho^K_\mu I^{N_K},
\sigma^1_\mu\ldots \sigma^M_\mu \right)}+
\diag{\left(\tilde x^1_\mu,\ldots \tilde x^K_\mu,
0\ldots 0\right)}\nn\\
&&\qquad+\sum_a \tau^a_\mu \bar t^a.
\label{SUN-2.1}\eea
Here $I^N$ denotes the $N\times N$ Identity matrix,  $\tilde x_\mu^k$ lies
in the Lie algebra $su(N_k)$ with $\tilde x_1^k$
diagonal, $M$ is given by
\beq M=N-\sum_{k=1}^K N_k \label{SUN-2.1.5}\eeq
and the tracelessness condition is 
\beq 0=\sum_{k=1}^M\sigma^k_\mu+\sum_{k=1}^K N_k\rho^k_\mu.\label{SUN-2.3}\eeq
Note that the ordering of $\rho$s and $\sigma$s in \ref{SUN-2.1} is not significant; for example we could chose to take $x_1$ so that the elements are in
decreasing order down the diagonal and then degenerate and non-degenerate 
eigenvalues would be all mixed up in general.
If $\tilde x_1^{1,\ldots K}=0$ then the eigenvalues of $x_1$ are exactly
degenerate; we have $K$ blocks of eigenvalues $\rho_1^{1,\ldots K}$ with degeneracy
$N_k$ and singleton eigenvalues $\sigma_1^{1,\ldots M}$. When the exact degeneracy is relaxed slightly we have blocks
of eigenvalues $\lambda^k_m, m=1,\ldots N_k$ with each block having
central value
\beq \frac{1}{N_k}\sum_{m=1}^{N_k}\lambda^k_m=\rho_1^k.\label{SUN-2.4}\eeq 
together with the singleton eigenvalues $\lambda^{(K+j)}_1=\sigma_1^j$, $j=1,\ldots M$.

 For each sub-algebra $H$ there is a contribution $\C I^{H}_{D,SU(N)}(R)$
to $\C I^{deg}_{D,SU(N)}(R)$. This comes from the integration region $\C P$
where  the $\rho_1^k, \sigma_1^j, \tilde x_1^k$
are  such that
 the 
eigenvalues of $x_1$ satisfy
\bea \C P: \quad\absval{\lambda^k_m-\lambda^{k'}_{m'}}&>&\epsilon,\quad k\ne k'
,\: m=1,\ldots N_k,\: m'=1,\ldots N_{k'}\nn\\
\absval{\lambda^k_m-\lambda^{k}_{m+1}}&\le &\epsilon, \quad m=1,\ldots N_k-1.
\label{SUN-2.5}
\eea
 We note that any sequence of eigenvalues can be arranged in the 
manner implied by \ref{SUN-2.5} for some $ H$. Therefore by considering 
the $\C I^{H}_{D,SU(N)}(R)$ for all possible $H$ we exhaust all possible
nearly degenerate eigenvalue configurations for $x_1$. That is to say
the degenerate term in $\C I_{D,SU(N)}(R)$ is now the sum of
contributions from  all
the possible $H$s 
\beq  \C I^{deg}_{D,SU(N)}(R)=\sum_H\C I^{H}_{D,SU(N)}(R).\label{SUN-2}\eeq

Now we bound the $\C I^{H}_{D,SU(N)}(R)$. 
The Vandermonde determinant for $x_1$ is easily  bounded from above through
\bea \Delta &=&\prod_{k>k', m,m'} (\lambda^k_m-\lambda^{k'}_{m'})^2
\prod_{k, m>m'}(\lambda^k_m-\lambda^{k}_{m'})^2\nn\\
&<&2^{n_H}\prod_{k, m>m'}(\lambda^k_m-\lambda^{k}_{m'})^2\nn\\
n_H&=&N(N-1)-\sum_{k=1}^K N_k(N_k-1) \label{SUN-4}\eea
where we have used the fact that none of the eigenvalues can have magnitude
 more
than 1 on account of the constraint \ref{P4}. Of course this bound is simply
 the product of the Vandermonde determinants  for the constituent
$su(N_k)$ factors of the sub-algebra $H$. 
Now we note
 that the $\tau_\mu^a$,  which are 
those off-diagonal elements of $x_{2,\ldots D}$
 that do not lie in
$H$, are constrained by \ref{P9} and, following our procedure in the $SU(3)$
case, we integrate them out to get 
\bea \lefteqn{\C I^{H}_{D,SU(N)}(R) <  C_1 R^{-(2-\eta)((D-1)n_H-2)}
\int_{\C P}\prod_{\mu=1}^D\left( \prod_{k=1}^K d\rho^k_\mu \prod_{j=1}^{M-1}
 d\sigma^j_\mu    \right)}
\nn\\&&\times \prod_{k=1}^K\left( \int_{\widetilde{\C R^k}} \prod_{\nu=1}^D d\tilde x^k_\nu\right)
\thetafn{\left(1-\sum_{k=1}^K\left(\Tr{ \tilde x^k_\mu\tilde x^k_\mu}+
N_k\rho^k_\mu\rho^k_\mu\right)-\sum_{j=1}^M\sigma^j_\mu\sigma^j_\mu  \right)}\nn\\
&&\times
\thetafn{\left(\sum_{k=1}^K\Tr{ \tilde x^k_\mu\tilde x^k_\mu}+
N_k\rho^k_\mu\rho^k_\mu+\sum_{j=1}^M\sigma^j_\mu\sigma^j_\mu -
\left(1-n_H\left(\epsilon R^{(2-\eta)}\right)^{-2}\right)\right)
}
\label{SUN-5}\eea
 where $\sigma^M_\mu$ is given by \ref{SUN-2.3},
the region
$\widetilde{\C R^k}$ is defined by
\beq \absval{\comm{\tilde x^k_\mu}{\tilde x^k_\nu}_{ij}}<R^{-(2-\eta)},\quad
\forall{\mu,\nu,i,j}.\label{SUN-6}\eeq

The right hand side of \ref{SUN-5} is now bounded above by dropping 
the $\C P$ constraint  and integrating out the $\rho_\mu^k $ and
$\sigma_\mu^j $ which leaves us with 
\bea \C I^{H}_{D,SU(N)}(R)& < & C_2 R^{-(2-\eta)(D-1)n_H}
\prod_{k=1}^K\left( \int_{\widetilde{\C R^k}} \prod_{\nu=1}^D d\tilde x^k_\nu\right)\thetafn{\left(1-\sum_{k=1}^K\Tr{ \tilde x^k_\mu\tilde x^k_\mu}\right)}\nn\\
& < & C_2 R^{-(2-\eta)(D-1)n_H}\prod_{k=1}^K \C F_{D,SU(N_k)}\label{SUN-7}\eea
where
\bea\C F_{D,SU(N_k)}= \left(\int_{\widetilde{\C R^k}} \prod_{\nu=1}^D d\tilde x^k_\nu\right)\thetafn{\left(1-\Tr{ \tilde x^k_\mu\tilde x^k_\mu}\right)}.
\label{SUN-8}\eea
We note in passing that if $H$ is empty then \ref{SUN-7} simply reduces
to the generic case \ref{P10}.
We now repeat the steps \ref{SU3-7} to \ref{SU3-9} to find that
\beq \C F_{D,SU(N_k)}=R^{-(2-\eta)D(N_k^2-1)/2}
\int_0^R u^{D(N_k^2-1)(1-\eta/2)-1}\C I_{D,SU(N_k)}(u)\,du\label{SUN-9}\eeq
The final step is by induction on $N$:
\begin{enumerate}
\item  $\C F_{D,SU(2)}$ is given in   
\ref{SU3-9}.
\item From our results for $SU(3)$ \ref{SU3-10} and 
\ref{SU3-11} we deduce that
\bea   \C F_{D,SU(3)}&<& C_3 R^{-4(2-\eta)D},\quad  D\ge 4\nn\\
	&&C_4 R^{-4(2-\eta)D}\log R,\quad D=3\label{SUN-10}\eea
\item For gauge group $SU(4)$ the possible sub-algebras, $H$, are
$su(2)$, $su(2)\oplus su(2)$ and $su(3)$. For $D=3$ we find that
\beq \C I_{3,SU(4)}(R)< C_5 R^{-48}\log R\label{SUN-11}\eeq
and hence the integral for $\C Z_{3,SU(4)}$ converges.  For 
$D\ge 4$ it is simple to check that $\C Z_{D,SU(4)}$ converges 
and the dominant term in $\C I_{D,SU(4)}(R)$  comes
from $H=su(3)$ (we will give a general formula for the behaviour
of $\C I_{D,SU(N)}(R)$ below). 
\item We now assume that convergence is established for $D\ge 3$ for
$N< N^*$.  As a consequence  we have that
\beq  \C F_{D,SU(N)}<C R^{-(2-\eta)D(N^2-1)/2}, \quad D\ge 3, \hbox{~and~}
3<N< N^*\label{SUN-12}\eeq 
together with the slightly different bounds \ref{SU3-9} and \ref{SUN-10} for $N=2$ and 3.
This is enough information to bound $\C I^H_{D,SU(N^*)}(R)$ using \ref{SUN-7}
 for
 any sub-algebra  $H$
\beq \C I^H_{D,SU(N^*)}(R)< C_6 R^{-(2-\eta)((D-1)n^*_H+\half D\sum_k
(N_k^2-1))}\left[R^{n_2\delta_{D,3}}(\log R)^{n_2\delta_{D,3}+
n_3\delta_{D,4}}\right]\label{SUN-13}\eeq 
where $n_{2,3}$ denotes the number of $su(2)$ and $ su(3)$ factors
respectively in $H$.
It is straightforward to check that for $D\ge 3$ the slowest decay at large $R$ occurs
when $H=su(N-1)$; basically this minimises the number of $R^{-2}$ factors 
coming from off-diagonal elements not in the sub-algebra
and maximises the number of $R^{-1}$ factors coming from elements in the  
sub-algebra.  Thus for $D\ge 3$ and $N\ge 4$ we have
\beq \C I^{deg}_{D,SU(N)}(R)< C R^{-(4(D-1)(N-1)+DN(N-2))}(\log R)^{
\delta_{N,4}\delta_{D,3}}\label{SUN-14}\eeq 
Applying the criterion \ref{P6.1} shows
that both generic \ref{P10} and degenerate terms make a finite contribution
to $\C Z_{D,SU(N)}$ for $D\ge 3$, $N\ge 4$.

\end{enumerate}
This completes our proof.

\section{Discussion}
\subsection{Correlation Functions}

We can extend the definition of the partition function \ref{P1} to
correlation functions so that 
\beq \expect{.}=\prod_{\mu=1}^D\int_{-\infty}^\infty dX_\mu(.)
\exp\left(\sum_{\mu>\nu}
\Tr \comm{X_\mu}{X_\nu}^2\right) \label{D1}\eeq
where $(.)$ represents some kind of 
 product of the $X_\mu$ with $P$ factors.
 Making the 
change of variables \ref{P4},  realising that the absolute value of the
corresponding product of the $x_\mu$ must be bounded by a constant, and using 
\ref{SUN-14} we get that
\beq \absval{\expect{.}}<C_8\int_0^RdR\,R^{-2N(D-2)+3D-5}\,R^P.\label{D2}\eeq
Thus correlators with fewer than
\beq P_c=2N(D-2)-3D+4\label{D3}\eeq
factors are guaranteed to be finite; of course correlators with more factors
than this \emph{may} be finite but then they must have some special
property so that the leading divergences cancel.
 The authors of \cite{Krauth:1999qw}
``guessed'' on the basis of reasonable arguments
that the eigenvalue density $\rho(\lambda)$ for $X_\mu$
behaves like
\beq \rho(\lambda)\sim const\,\lambda^{-2N(D-2)+3D-5}\label{D4}\eeq
at large $\lambda$. This is completely consistent with our results.

\subsection{Supersymmetric Integrals}

The supersymmetric partition functions are given by 
\beq \C Z^{SS}_{D,G}=\prod_{\mu=1}^D\int_{-\infty}^\infty dX_\mu
\,\C P_{D,G}(X_\mu)\,	\exp\left(\sum_{\mu>\nu}
\Tr \comm{X_\mu}{X_\nu}^2\right) \label{SS1}\eeq
where the Pfaffian $\C P_{D,G}$ arises from integrating out the fermionic
degrees of freedom and is a homogeneous polynomial of degree
$(D-2)(N^2-1)$. We can of course regard $\C Z^{SS}_{D,G}$ as being a
correlation function in the bosonic theory and apply the considerations
of section 6.1 to it for $P=(D-2)(N^2-1)$;
 we find immediately that all the supersymmetric
partition functions are naively divergent. However the Pfaffian contains 
many terms with different signs and we expect many cancellations.
The simplest example is for $SU(2)$ where the Pfaffians are known explicitly
\cite{Krauth:1998xh}  and (except for $D=3$ where $\C Z^{SS}_{3,SU(2)}=0$ because $\C P_{3,SU(2)}$
is an odd function)  can be expressed as sums of powers of commutators
$\comm{X_\mu}{X_\nu}$. This is particularly convenient with our method 
because the rescaling in \ref{P4} followed by the restriction to the region
$\C R_1$ in \ref{P7} means that we can bound $\C Z^{SS}_{D,SU(2)}$ simply
by setting all commutators to a constant. It follows that if the bosonic
partition function converges so does the supersymmetric one; thus the
$D=6$ and 10 partition functions are convergent but $D=4$ is marginal and
we  would need to work harder (in fact it is known to converge). The situation
with bigger $N$ is more 
complicated mainly because relatively little is known about the
Pfaffians and we will return to this problem in a separate paper.

\subsection{Other Gauge Groups}

In this paper we have concentrated on $SU(N)$ gauge groups. However all
the results we obtain depend in a well defined way on simple
group theoretical  properties such as the order, rank and sub-algebras
of Lie algebras.  It is therefore tempting to suppose that expressed
in this form our results would carry over directly to any Lie group.
However some of the steps we have made such as the diagonalization
of $x_1$ and the inductive argument in section 5 
do depend on the group being $SU(N)$ and we will deal 
with the other groups in a separate paper.

\acknowledgments
 JFW acknowledge valuable discussions with Bergfinnur Durhuus,
 Thordur Jonsson, George Savvidy and Konstantin Savvidy.
PA acknowledges a PPARC studentship.

\appendix
\section{Divergent Matrix Integrals}
The $D=2$ integral
\beq \C Z_{2,G}=\int_{-\infty}^\infty dX_1 dX_2\exp\left(
\Tr \comm{X_1}{X_2}^2\right) \label{AA1}\eeq
 is divergent for all $SU(N)$. This is easily seen by 
diagonalizing $X_1$; the integrand then does not depend upon
the diagonal elements of $X_2$ and so the integral over them diverges.

Some other low $N$ and low  $D$ integrals are  divergent. To see this 
we go back to \ref{P1}, diagonalize $X_1$, and separate out the diagonal
elements for $\nu>1$,
\beq X_\nu=\hbox{diag}(\lambda_{\nu 1},\ldots \lambda_{\nu N})
+X^\perp_\nu.\label{AA2}\eeq
We then change variables from the $X^\perp_\nu$ to
$(D-1)(g-l)$ dimensional polar coordinates with radial variable
$r$ and angular variables $\{\theta_i\}$. The integral over all the diagonal
elements is gaussian so we do it and are left with
\beq \C Z_{D,G}=\int_0^\infty r^{D(g-2l)-2g+2l-1} dr\int d\Omega F_1(\{\theta_i\})
\exp(-r^4F_2(\{\theta_i\}))\label{AA3}\eeq
where $F_1$ and $F_2$ are horrible but positive semi-definite functions
and $\Omega$ is the $(D-1)(g-l)$ dimensional solid angle. We see immediately
that the integral over $r$ diverges at $r=0$  if
\beq D \le \frac{2(g-l)}{g-2l}\label{AA4}\eeq
so we deduce that the $D=3,4$ integrals diverge for $SU(2)$ and that
the $D=3$ integral diverges for $SU(3)$.

\bibliographystyle{JHEP}
\bibliography{YMrefs}

\providecommand{\href}[2]{#2}\begingroup\raggedright\begin{thebibliography}{10}

\bibitem{Banks:1997vh}
T.~Banks, W.~Fischler, S.~H. Shenker, and L.~Susskind, {\it {M Theory As A
  Matrix Model: A Conjecture}},  {\em Phys. Rev.} {\bf D55} (1997) 5112--5128,
  [\href{http://xxx.lanl.gov/abs/hep-th/9610043}{{\tt hep-th/9610043}}].

\bibitem{Ishibashi:1997xs}
N.~Ishibashi, H.~Kawai, Y.~Kitazawa, and A.~Tsuchiya, {\it {A Large-N Reduced
  Model as Superstring}},  {\em Nucl. Phys.} {\bf B498} (1997) 467--491,
  [\href{http://xxx.lanl.gov/abs/hep-th/9612115}{{\tt hep-th/9612115}}].

\bibitem{Taylor:1999qk}
{Washington Taylor IV}, {\it {The M(atrix) model of M-theory}},
  \href{http://xxx.lanl.gov/abs/hep-th/0002016}{{\tt hep-th/0002016}}.

\bibitem{Aoki:1998bq}
H.~Aoki, S.~Iso, H.~Kawai, Y.~Kitazawa, A.~Tsuchiya, and T.~Tada, {\it Iib
  matrix model},  {\em Prog. Theor. Phys. Suppl.} {\bf 134} (1999) 47,
  [\href{http://xxx.lanl.gov/abs/hep-th/9908038}{{\tt hep-th/9908038}}].

\bibitem{Sethi:1997pa}
S.~Sethi and M.~Stern, {\it {D-Brane Bound States Redux}},  {\em Commun. Math.
  Phys.} {\bf 194} (1998) 675,
  [\href{http://xxx.lanl.gov/abs/hep-th/9705046}{{\tt hep-th/9705046}}].

\bibitem{Yi:1997eg}
P.~Yi, {\it {Witten Index and Threshold Bound States of D-Branes}},  {\em Nucl.
  Phys.} {\bf B505} (1997) 307,
  [\href{http://xxx.lanl.gov/abs/hep-th/9704098}{{\tt hep-th/9704098}}].

\bibitem{Kac:1999av}
V.~G. Kac and A.~V. Smilga, {\it {Normalized Vacuum States in N = 4
  Supersymmetric Yang-Mills Quantum Mechanics with Any Gauge Group}},  {\em
  Nucl. Phys.} {\bf B571} (2000) 515,
  [\href{http://xxx.lanl.gov/abs/hep-th/9908096}{{\tt hep-th/9908096}}].

\bibitem{Hoppe:1999xg}
J.~Hoppe, V.~Kazakov, and I.~K. Kostov, {\it {Dimensionally Reduced SYM(4) as
  Solvable Matrix Quantum Mechanics}},  {\em Nucl. Phys.} {\bf B571} (2000)
  479, [\href{http://xxx.lanl.gov/abs/hep-th/9907058}{{\tt hep-th/9907058}}].

\bibitem{Staudacher:2000gx}
M.~Staudacher, {\it {Bulk Witten Indices and the Number of Normalizable Ground
  States in Supersymmetric Quantum Mechanics of Orthogonal, Symplectic and
  Exceptional Groups}},  {\em Phys. Lett.} {\bf B488} (2000) 194,
  [\href{http://xxx.lanl.gov/abs/hep-th/0006234}{{\tt hep-th/0006234}}].

\bibitem{Moore:1998et}
G.~Moore, N.~Nekrasov, and S.~Shatashvili, {\it D-particle bound states and
  generalized instantons},  {\em Commun. Math. Phys.} {\bf 209} (2000) 77,
  [\href{http://xxx.lanl.gov/abs/hep-th/9803265}{{\tt hep-th/9803265}}].

\bibitem{Suyama:1998ig}
T.~Suyama and A.~Tsuchiya, {\it Exact results in n(c) = 2 iib matrix model},
  {\em Prog. Theor. Phys.} {\bf 99} (1998) 321--325,
  [\href{http://xxx.lanl.gov/abs/hep-th/9711073}{{\tt hep-th/9711073}}].

\bibitem{Krauth:1998xh}
W.~Krauth, H.~Nicolai, and M.~Staudacher, {\it {Monte Carlo Approach to
  M-theory}},  {\em Phys. Lett.} {\bf B431} (1998) 31--41,
  [\href{http://xxx.lanl.gov/abs/hep-th/9803117}{{\tt hep-th/9803117}}].

\bibitem{Krauth:1998yu}
W.~Krauth and M.~Staudacher, {\it {Finite Yang-Mills Integrals}},  {\em Phys.
  Lett.} {\bf B435} (1998) 350,
  [\href{http://xxx.lanl.gov/abs/hep-th/9804199}{{\tt hep-th/9804199}}].

\bibitem{Krauth:1999qw}
W.~Krauth and M.~Staudacher, {\it Eigenvalue distributions in yang-mills
  integrals},  {\em Phys. Lett.} {\bf B453} (1999) 253--257,
  [\href{http://xxx.lanl.gov/abs/hep-th/9902113}{{\tt hep-th/9902113}}].

\bibitem{Ambjorn:2000bf}
J.~Ambjorn, K.~N. Anagnostopoulos, W.~Bietenholz, T.~Hotta, and J.~Nishimura,
  {\it Large n dynamics of dimensionally reduced 4d su(n) super yang-mills
  theory},  {\em JHEP} {\bf 07} (2000) 013,
  [\href{http://xxx.lanl.gov/abs/hep-th/0003208}{{\tt hep-th/0003208}}].

\bibitem{Hotta:1998en}
T.~Hotta, J.~Nishimura, and A.~Tsuchiya, {\it Dynamical aspects of large n
  reduced models},  {\em Nucl. Phys.} {\bf B545} (1999) 543--575,
  [\href{http://xxx.lanl.gov/abs/hep-th/9811220}{{\tt hep-th/9811220}}].

\bibitem{Krauth:2000bv}
W.~Krauth and M.~Staudacher, {\it {Yang-Mills Integrals for Orthogonal,
  Symplectic and Exceptional Groups}},  {\em Nucl. Phys.} {\bf B584} (2000)
  641, [\href{http://xxx.lanl.gov/abs/hep-th/0004076}{{\tt hep-th/0004076}}].

\end{thebibliography}\endgroup

\end{document}